%% file: gpu-BG.tex
\documentclass[11pt,epsf]{article}
 \usepackage{amsmath}
 \usepackage{amssymb}
 \usepackage{graphicx}
 \usepackage[merge,numbers,compress]{natbib}
 \usepackage[T1]{fontenc}
 \usepackage{booktabs}
 \usepackage{xcolor} 
 \usepackage{xspace}
 \usepackage{dcolumn}
 \usepackage{placeins}
 \usepackage[colorlinks=true,citecolor=blue!50!black,linkcolor=black]{hyperref}
 \usepackage{caption}
 \usepackage[position=top,skip=-2pt,justification=centering,margin=15pt,singlelinecheck=false]{subcaption}
 \usepackage{multirow}
 \graphicspath{{./figures/}}

\usepackage[colorinlistoftodos]{todonotes}

\input{macros}

\begin{document}

\title{\hfill ~\\[-30mm]
\phantom{h} \hfill\mbox{\small CERN-TH-2025-017, FR-PHENO-2025-001}
\\[1cm]
\vspace{10mm} \textbf{Accelerating Berends-Giele recursion for gluons
  \\ in arbitrary dimensions over finite fields}}

\date{}
\author{
Juan Cruz-Martinez$^{1\,}$\footnote{E-mail:
  \texttt{juan.cruz.martinez@cern.ch}},
Giuseppe De Laurentis$^{2\,}$\footnote{E-mail:
  \texttt{giuseppe.delaurentis@ed.ac.uk}},
Mathieu Pellen$^{3\,}$\footnote{E-mail:
  \texttt{mathieu.pellen@physik.uni-freiburg.de}}
\\[9mm]
{\small\it $^1$ CERN, Theoretical Physics Department,} \\
{\small\it CH-1211 Geneva 23, Switzerland}\\[3mm]
{\small\it $^2$ Higgs Centre for Theoretical Physics, University of Edinburgh, } \\
{\small\it Edinburgh, EH9 3FD, United Kingdom}\\[3mm]
{\small\it $^3$ Universit\"at Freiburg, Physikalisches Institut,} \\
{\small\it Hermann-Herder-Str. 3, 79104 Freiburg, Germany}\\[3mm]
}
\maketitle

\begin{abstract}
\noindent
This work provides a proof of concept for the computation of pure
gluonic amplitudes in quantum chromodynamics (QCD) on graphics
processing units (GPUs). The implementation relies on the
Berends-Giele recursion algorithm and, for the first time on a GPU,
enables the numerical computation of amplitudes in an arbitrary number
of space-time dimensions and over finite fields. This demonstrates the
advantages of hardware acceleration, not only for the computation of
tree-level amplitudes for real-radiation processes in four dimensions
over complex numbers but also for generating loop integrands for
virtual corrections in $d$ dimensions over finite fields. The
associated computer program is publicly available.
\end{abstract}
\thispagestyle{empty}
\setcounter{page}{1}

\newpage

\noindent\rule{\linewidth}{1pt}
\vspace{-2em}
\tableofcontents
\vspace{0.5em}
\noindent\rule{\linewidth}{1pt}

\section{Introduction}

The precision phenomenology program at CERN's Large Hadron Collider
(LHC) requires very precise and accurate theoretical calculations in
order to match the unprecedented precision achieved by experimental
measurements~\cite{Huss:2022ful}. Theoretical prediction, mainly obtained through Monte
Carlo techniques, are very intensive computationally and their cost is
expected to grow in the future as we move towards the high-luminosity
phase of the LHC, when even more accurate predictions will be
necessary
\cite{Buckley:2019wov,HSFPhysicsEventGeneratorWG:2020gxw,Campana:2022etc,FebresCordero:2022psq}.
In addition, this increasing demand for computational resources is
affecting all aspects of theoretical high-energy physics, ranging from
parton distribution function (PDF) determination, to loop calculations
and Monte Carlo (MC) integration. To remedy this problem, in addition
to developing custom solutions, the use of graphics processing units
(GPUs) has been found to be a particularly attractive solution given
the high parallelisability of many particle-physics problems.

In recent years, the rapid development of GPUs, driven primarily by
the growing demand for artificial intelligence (AI) applications, has
transformed them into general-purpose graphics-processing units
(GPGPUs), extending their use beyond traditional graphics rendering to
encompass high-performance scientific computing. While central
processing units (CPUs) have traditionally been optimised for
high-speed execution of single tasks, making them well-suited to
sequential processing, GPUs are designed to handle large sets of
similar tasks in parallel, prioritising massive throughput over
per-task execution speed. Thus, the key difference between optimising
code for GPUs and CPUs lies in the shift to maximising data throughput
by exposing as much concurrency as possible. In turn, this can lead to
trade-offs in performance when the calculation has to be executed
sequentially. However, in today's era of multi-core processors and
large computing clusters, many of the optimisation techniques
targeting hardware accelerators can similarly be advantageous on
traditional CPUs. Currently, most GPGPU libraries focus on float or
double precision types, which the devices can support natively. More
complex numeric types require custom implementations, resulting in
performance penalties and the inability to leverage general-purpose
libraries. These limitations can affect development efficiency and
flexibility.

Particle physics has not remained indifferent to these developments.
While the most-used software in particle physics phenomenology are
still CPU-based, the interest in GPGPU has translated into libraries
for device-agnostic Monte Carlo
integration~\cite{Borowka:2018goh,Carrazza:2020rdn,torchquad:2021},
for PDF determination~\cite{NNPDF:2021uiq,Cruz-Martinez:2024wiu},
interpolation~\cite{Carrazza:2020qwu}, parton shower~\cite{Seymour:2024fmq}, and amplitude generation and
evaluation~\cite{Bothmann:2021nch,Heinrich:2023til}.  Some of these
efforts have culminated in the recent years on full device-agnostic
event
generators~\cite{Carrazza:2021gpx,Carrazza:2022jbl,Bothmann:2023gew}
targeting leading-order (LO) calculations. Comparative studies
presented in the referred work show gains that can go up to a factor
of two orders of magnitude for some exemplary cases.  It is also worth
mentioning efforts in \madgraph to migrate to a fully GPU
implementation, with preliminary results shown in
Refs.~\cite{Valassi:2021ljk,Valassi:2022dkc,Valassi:2023yud,Wettersten:2023ekm,Hageboeck:2023blb}.

If we restrict ourselves to fixed orders calculation, only LO
generators are available, although a proof-of-concept calculation at
next-to-leading order (NLO) QCD on a GPU was presented in
Ref.~\cite{Carrazza:2020qwu} showing that these devices were capable
of handling more complicated calculations beyond massive
parallelisation.  This last example, however, was obtained with
hard-coded matrix elements with all the limitations that it implies.
As of now, there exists efficient methods to obtain fully general numerical tree and one-loop amplitudes in the Standard Model and beyond based on recursion
techniques~\cite{Berends:1987me,vanHameren:2009vq}.  These algorithms
have lead to the implementation of efficient numerical codes running
on CPUs~\cite{Kanaki:2000ey,Cascioli:2011va,Actis:2016mpe}.  It is
therefore of prime importance to design new computer programs able to
handle these algorithms on GPUs.  The first proof of concept in this
direction was released more than 10 years ago
\cite{Giele:2010ks,Hagiwara:2009cy} but was restricted to massless QCD
with only one public program limited to specific QCD
processes~\cite{Hagiwara:2009cy,Hagiwara:2010oca}.  The recent efforts
in the context of
\madgraph~\cite{Carrazza:2021gpx,Carrazza:2022jbl,Valassi:2021ljk,Valassi:2022dkc,Valassi:2023yud,Wettersten:2023ekm,Hageboeck:2023blb}
and {\sc Sherpa}~\cite{Bothmann:2021nch,Bothmann:2023gew} points to a
future in which GPUs will be regularly used in event generation.
However, presently these efforts are limited to leading-order event
generator and no matrix-element provider targeting high-accuracy
calculations is available.

In particular, while most of the effort has been focusing on the
development of Monte Carlo aspects where the gains expected from GPU
implementations are the most obvious, the advantages of using GPU for
loop calculations can also be critical. Such an example has already
been provided in Ref.~\cite{Li:2015foa,Borowka:2018goh} where loop
integrals are computed numerically using GPU-based integration
routines.
In this work, we explore a different potential application of GPU
acceleration in the context of multi-loop amplitude computations,
namely the computation of tree amplitudes required to construct the
integrand in the numerical unitarity method, \emph{e.g.}\ as
implemented on CPUs in \textsc{Caravel} \cite{Abreu:2020xvt}. The
challenge, so far not tackled on GPUs, is that trees in this context
have to be computed over finite fields, to avoid numerical
instabilities, and in arbitrary $d$ dimensions, to capture terms
naively missed by $d=4$ cuts. A large number of finite field samples
is then used for either analytic reconstruction
\cite{vonManteuffel:2014ixa,Peraro:2016wsq} or rational-number
reconstruction at rational kinematic points (see \emph{e.g.}\ Refs.~\cite{Hartanto:2019uvl,Badger:2024dxo}).
This part of the calculation can therefore be particularly
computationally intensive, and it could benefit from massive
parallelisation on GPUs.

For these reasons, we have developed the first public program for the
computation of tree-level amplitudes in arbitrary dimensions using different numeric types.
On the CPU, the program supports arbitrary precision floats and finite fields,
while on the GPU it supports double precision floats and finite fields.
In this study we rely on integer arithmetic capabilities of GPUs.
This proof of concept is for now restricted to purely gluonic amplitudes
but it is foreseen to be extended to the full Standard Model in the
future. The code is publicly available at
\begin{center}
 \url{https://github.com/Amps-GPU/BG-Trees}
\end{center}
and it can be readily used. We provide an \href{https://github.com/Amps-GPU/BG-Trees/blob/main/example/Example.ipynb}{example jupyter notebook} within the repository.

The organisation of the article is as follows.
Section~\ref{sec:method} introduces the general motivation of the work, the formalism used as well as the conventions.
In Section~\ref{sec:gpu}, the general strategy is presented along with the checks that have been performed and a benchmarking exercise.
Section~\ref{sec:concl} contains a summary of the work as well as concluding remarks.

\section{Use cases and method overview}
\label{sec:method}

In this section, we motivate the importance of tree-level amplitudes and their implementations in GPU.
After this, the Berends-Giele recursion used in the present work is briefly introduced.
Finally, the notations and conventions that we have adopted for our implementation in $d$ dimension are presented.

\subsection{Context}

While LO accuracy is no longer the standard for
particle-physics phenomenology, tree-level amplitudes continue to be
of critical interest.  For example, in
Refs.~\cite{Hoche:2019flt,Bothmann:2022thx}, it has been found that,
for typical theoretical predictions, a fair share of the total amount
of CPU time is spent in evaluating tree-level matrix elements.  Also,
for fixed-order predictions, tree-level amplitudes are needed for
subtraction terms or real radiations and thus play a crucial role, in
particular regarding to their numerical stability.

In additions to these well known use cases, the generalised
  unitarity method~\cite{Bern:1994zx, Bern:1994cg, Bern:1997sc,
    Britto:2004nc} in its numerical \cite{Ellis:2007br, Giele:2008ve,
    Berger:2008sj} and multi-loop \cite{Abreu:2017idw, Abreu:2017xsl}
  formulation,\footnote{For earlier analytical developments and
  applications, see Refs.~\cite{Gluza:2010ws, Badger:2012dp,
    Badger:2013gxa, Ita:2015tya, Badger:2015lda} and references
  therein.} also requires the fast and reliable evaluation of tree
  amplitudes, usually through Berends-Giele recursion
  \cite{Berends:1987me}.  In the following, we summarise this method
to compute multi-loop amplitudes, as implemented in the computer code
{\sc Caravel}~\cite{Abreu:2020xvt}.

In general, the non-colour part of an amplitude can be written as
\begin{align}
 \mathcal{A}^{(L)} = \sum_{\Gamma \in \Delta} \sum_{i \in M_\Gamma} c_{\Gamma, i} \mathcal{I}_{\Gamma, i} ,
\end{align}
where the set $\Delta$ contains all propagator structures $\Gamma$ while $M_\Gamma$ is the full set of master integrals associated to $\Gamma$.
The master integrals are thus $\mathcal{I}_{\Gamma, i}$ while the coefficients $c_{\Gamma, i}$ are functions of the kinematic invariants.
The corresponding integrand, with $\ell_l$ the set of $L$ loop momenta, can be further parametrised as
\begin{align}
\label{eq:integrand}
 \mathcal{A}^{(L)}(\ell_l) = \sum_{\Gamma \in \Delta} \sum_{k \in Q_\Gamma} c_{\Gamma, k} 
 \frac{m_{\Gamma,k}(\ell_l)}{\prod_{j \in P_\Gamma} \rho_j(\ell_1)} .
\end{align}
The set $P_\Gamma$ contains all inverse propagators $\rho_j$ for the diagram $\Gamma$ (or equivalently the propagator structure) while the $Q_\Gamma$ labels all the possible integrand $m_{\Gamma,k}$.

In order to compute the coefficients $c_{\Gamma, k}$ of Eq.~\eqref{eq:integrand}, 
the properties of loop integrands for on-shell configurations $\ell^\Gamma_l$ of the loop momenta are exploited.
These configurations are defined for a propagator structure $P_\Gamma$ as
\begin{align}
 \rho(\ell^\Gamma_l) = 0 \quad \textrm{iff} \quad j \in P_\Gamma .
\end{align}
In this case, the leading pole of Eq.~\eqref{eq:integrand} reads
\begin{align}
 \sum_{\rm states} \prod_{i \in T_\Gamma} \mathcal{A}_i^{\rm tree}
 \left(\ell^\Gamma_l \right) = \sum_{\Gamma' \ge \Gamma,\;  k \in Q_{\Gamma'}}
 \frac{c_{\Gamma',k} m_{\Gamma',k}\left(\ell^\Gamma_l \right)}{\prod_{j\in(P_{\Gamma'}/P_{\Gamma'})} \rho\left(\ell^\Gamma_l \right)} .
\end{align}
In this case, $T_\Gamma$ is the set of all tree-level amplitudes corresponding to the
vertices of $\Gamma$ while the state sum runs over all $d_s$-dimensional particle states appearing in the loop.
The diagrams $\Gamma'$ have an equal or larger number of propagators than $\Gamma$.
Using this cut equation to determine the coefficients $c_{\Gamma, k}$ therefore requires the multiple evaluation (in order to generate a set of linear equations) of the tree amplitudes $\mathcal{A}_i^{\rm tree}$ at many phase-space points.
This step represents a substantial part of the total computational time required for obtaining the loop amplitude.
It means that having a program offering a fast evaluation of tree-level amplitudes in arbitrary dimensions is particularly valuable.
This is one of the main motivations for the present work.

\subsection{Berends-Giele recursion}

In QCD, any gluon amplitude can be cast in the following form
\begin{align}
\mathcal{M}_n(1,\ldots,n) = \sum_{P(1,\ldots,n-1)} \textrm{Tr}(T^{a_1} \ldots T^{a_n}) \, \mathcal{C}(k_1^{\lambda_1},\ldots,k_n^{\lambda_n}),
\end{align}
where $\mathcal{C}$ is a function enclosing the information about the momenta and the helicities of the gluons.
The indices $a_i$ denote the colours of the $n$ gluons and $T^{a_i}$ are colour matrices in the fundamental representation.

The colour part of the amplitude can be obtained with analytical or numerical techniques.
On the other hand, the non-colour part \emph{i.e.}\ the function $\mathcal{C}$ can be obtained with various numerical or analytical techniques.
In the present work, we revert to the Berends-Giele (BG) recursion~\cite{Berends:1987me} which has been shown to be particularly efficient.
The general idea is that the kinematic part of the amplitudes can be efficiently built out of building blocks of lower multiplicities called \emph{currents}.

In particular, the matrix element for $n$ gluons can be built out of the $n-1$-gluon current by removing the propagator of the off-shell gluon, by contracting the current with the polarisation vector of the $n^{\rm th}$ gluon, and by enforcing momenta conservation.
The currents are then recursively defined as
\begin{align}
 J^\mu(k_1^{\lambda_1}, \ldots, k_n^{\lambda_n}) ={}& \frac{-\ri}{P^2_{1,n}} \bigg[ \sum^{n-1}_{i=1} V^{\mu\nu\rho}_3 (P_{1,i},P_{i+1,n}) J_\nu(k_1^{\lambda_1}, \ldots, k_i^{\lambda_i}) J_\rho(k_{i+1}^{\lambda_{i+1}}, \ldots, k_n^{\lambda_n}) \nonumber \\
 +{}& \sum^{n-2}_{i=1} \sum^{n-1}_{j=i+1} V^{\mu\nu\rho\sigma}_4 J_\nu(k_1^{\lambda_1}, \ldots, k_i^{\lambda_i}) J_\rho(k_{i+1}^{\lambda_{i+1}}, \ldots, k_j^{\lambda_j}) J_\sigma(k_{j+1}^{\lambda_{j+1}}, \ldots, k_n^{\lambda_n})\bigg] ,
\end{align}
where the first and second term account for the triple- and quartic-gluon vertices, respectively.
Furthermore,
\begin{align}
 V^{\mu\nu\rho}_3(P,Q) ={}& \ri \left( g^{\nu\rho} (P-Q)^\mu + 2 g^{\rho\mu} Q^\nu - 2 g^{\mu\nu} P^\rho \right), \nonumber \\
 V^{\mu\nu\rho\sigma}_4 ={}& \ri \left(2 g^{\mu\rho} g^{\nu\sigma} - g^{\mu\nu} g^{\rho\sigma} - g^{\mu\sigma} g^{\nu\rho} \right) , \nonumber \\
 P_{i,j} ={}& \sum_{l=i}^j k_l .
\end{align}

\subsection{Notations in $d$ dimensions}

Here the definitions and related conventions of the polarisation states are provided, taking the example of 6 dimensions.
In the case the polarisation is built for a massless 6-dimensional vector $k = \left\{ k^0, \ldots, k^5 \right\}$ (obeying therefore the relation $(k,k)=0$ ) and a 4-dimensional massless vector $\chi$ embedded in six dimensions as follow
\begin{align}
 \chi = \left\{ \chi^0, \ldots, \chi^3, 0, 0 \right\}, \quad \textrm{with} \quad (\chi,\chi)=0 .
\end{align}
The 4-dimensional part of the vector $k$ vector is defined as
\begin{align}
 k_{4d} = \left\{ k^0, \ldots, k^3, 0, 0 \right\} , \quad \textrm{with} \quad (k_{4d},k_{4d})=\mu^2 \neq 0 .
\end{align}
Hence, a massless four-dimensional vector can then be constructed in the following way
\begin{align}
 k_{4d}^\flat = k_{4d} - \frac{\mu^2}{2(\xi,k_{4d})} \chi.
\end{align}
Square roots are cumbersome when working in finite fields $\mathbb{F}_p$, since they
require a field extension $50\%$ of the time. However, they can be
avoided by multiplying $\epsilon$ by appropriate factors while dividing
out from the ``conjugate'' state $\epsilon^*$. In this case the two
states are no longer complex conjugate, like for spinors with complex
momenta. Nevertheless the completeness relation and orthonormality
conditions are still satisfied.
The polarisation vectors can be written as
\begin{alignat}{2}
 \epsilon_1 =&{} \left\{ \frac{\langle k_{4d}^\flat | \gamma^\mu | \chi ] }{ 2 \left[ k_{4d}^\flat , \chi\right] } ,0, 0 \right\}, & \quad \epsilon_2 &= \left\{ \frac{[ k_{4d}^\flat | \gamma^\mu | \chi \rangle }{ \langle k_{4d}^\flat , \chi\rangle } ,0, 0 \right\} , \nonumber \\
 \epsilon_3 =&{} \left\{ \frac{1}{\mu} k^\flat_{4d} - \frac{\mu}{2(\chi,k_{4d})} \chi ,0, 0 \right\}, &
 \quad \epsilon_4 &= \frac{1}{\mu}\left\{ 0,0,0,0,k_5,-k_4 \right\} .
\end{alignat}
By definition, the polarisation vectors are transverse $(k, \epsilon_i)=0$ and fulfil the following properties
\begin{align}
 (\epsilon_1,\epsilon_2) ={}& -1, \nonumber \\
 (\epsilon_3,\epsilon_3) ={}&  (\epsilon_4,\epsilon_4) = - 1,  \nonumber \\
 (\epsilon_1,\epsilon_1) ={}& (\epsilon_2,\epsilon_2) =  (\epsilon_3,\epsilon_4) =  (\epsilon_{1,2},\epsilon_{3,4}) = 0.
\end{align}
Note that we have adopted the conventions of Ref.~\cite{Gnendiger:2017pys} where the following expressions are provided in arbitrary dimensions.
In particular, the completeness relation reads
\begin{align}\label{eq:completess_relation_d_dim}
 \sum^4_{i=1} \epsilon_i^\mu \epsilon_i^{*\nu} = - g_{6d}^{\mu\nu} + \frac{k^\mu \eta^\nu + k^\nu \eta^\mu}{( k, \eta)} ,
\end{align}
where $\eta$ is the $6$-dimensional reference vector such that $(k,\eta)\neq0$ and $g_{6d}^{\mu\nu}$ is the mostly negative metric in $6$ dimensions.
This relation can be further split into a $4$-dimensional and two-dimensional part as
\begin{align}\label{eq:completess_relation_d_dim_split}
\sum^4_{i=1} \epsilon_i^\mu \epsilon_i^{*\nu} = \left( - g_{4d}^{\mu\nu} + \frac{k^\mu_{4d} k^\nu_{4d}}{\mu^2}\right) - \left( g_{2d}^{\mu\nu} + \frac{k^\mu_{2d} k^\nu_{2d}}{\mu^2} \right) ,
\end{align}
by taking the particular choice
\begin{align}
 \eta^\mu = k^\mu_{4d} - k^\mu_{2d} .
\end{align}

This construction of the polarisation states can be generalised to $d>6$ by introducing more polarisation states, where keeping the number of states equal to $d -
2$. We define the extra polarisation states beyond $d=6$ as, for example,
\begin{align}
\label{eq:extra-states5}
 \quad \epsilon^\mu_{5} ={}& \frac{1}{\mu_{x+1}\mu_{x+2}}\left\{ 0,0,0,0,k_4k_x, -\mu^2_{x+1},\vec 0\right\} \, ,
\end{align}
and more generally as
\begin{align}
  \quad \epsilon^\mu_{5\leq x \leq d-2} ={}& \frac{1}{\mu_{x+1}\mu_{x+2}}\left\{ 0,0,0,0,k_4k_x,\dots,k_{x-1}k_x, -\mu^2_{x+1},\vec 0\right\} \, ,
  \label{eq:extra-states}
\end{align}
where we define $\mu_{x}$ as
\begin{align}
\quad \mu^2_{x} = \sum_{i=4}^{x-1} k_i^2
\end{align}
meaning that $\mu^2 = \mu^2_{d}$.
The lower dots in Eq.~\eqref{eq:extra-states} denotes all terms of the form $k_i k_x $ with the index $i$ running from $5$ to $x-2$.
The expressions in
Eqs.~\eqref{eq:extra-states5} and \eqref{eq:extra-states} can be obtained by requiring that the
completeness relation and orthogonality conditions are satisfied while
solving for the polarisation components. It is also convenient to
require that $\epsilon^{\mu>x+1}_x=0$, so that lower dimensional
states are trivially embedded in higher dimensions.
Keeping the same convention to deal with the square roots, 
the polarisation vectors read
\begin{alignat}{2}
 \epsilon_1 =&{} \tilde\epsilon_2 = \left\{ \frac{\langle k_{4d}^\flat | \gamma^\mu | \chi ] }{ 2 \left[ k_{4d}^\flat , \chi\right] } ,0, 0 \right\}, & \quad \tilde\epsilon_1 &= \epsilon_2 = \left\{ \frac{[ k_{4d}^\flat | \gamma^\mu | \chi \rangle }{ \langle k_{4d}^\flat , \chi\rangle } ,0, 0 \right\} , \nonumber \\
 \epsilon_3 =&{} \left\{ k^\flat_{4d} - \frac{\mu^2}{2(\chi,k_{4d})} \chi ,0, 0 \right\}, & \quad  \tilde\epsilon_3 &= \frac{\epsilon_3}{\mu^2} , \nonumber \\
 \quad \epsilon_4 =&{} \left\{ 0,0,0,0,k_5,-k_4 \right\} , & \quad \tilde\epsilon_4 &= \frac{\epsilon_4}{\mu^2}.
\end{alignat}

It is worth pointing out that for loop integrals, only some of the external states will actually be $d$ dimensional while the others will be four dimensional.
Currently, four-dimensional states are trivially padded into $d$ dimensional states by appending zeroes.
Thus in this respect, the program treats all states, regardless of their actual dimension, on the same footing.
Having a specific recursion for the four-dimension case could be more efficient.
Investigating this aspect and potentially implementing a solution is left for future work.

\section{GPU acceleration}
\label{sec:gpu}
In this section we describe the specific GPU implementation.
In this work we achieve the GPU acceleration of $d$-dimensional amplitudes using finite fields through two strategies.
First, our focus is on future flexibility rather than solely performance.
To that end, we utilise the TensorFlow library~\cite{tensorflow2015-whitepaper}.
TensorFlow is a well-maintained library with a focus on Artificial Intelligence.
It provides many kernels for mathematical operations capable of running in GPUs (of different vendors) as well as CPUs.
Most importantly, it provides an easy interface for custom types, avoiding the need for specific implementations for each operation (at the cost of some overhead).
Note that while our focus is on finite fields the library can operate with arbitrary precision objects (through the \texttt{mpmath} library~\cite{mpmath}) as well as standard (\texttt{numpy}~\cite{harris2020array} or \texttt{tensorflow}~\cite{tensorflow2015-whitepaper}) complex numbers. As detailed below, only the finite-fields implementation has been explicitly optimised, while single and double real or complex numbers can run on GPU through tensorflow's standard capabilities.

When utilising finite field, there are several technical challenges to consider related to the representation of the number in the native supported types of the device.
We solve them by setting the default type of all integer arithmetic to 64-bits, while limiting the size of the finite field to $p < 2^{31}$, with a choice of $p$ such that the imaginary unit is contained in the field, $i \in \mathbb{F}_p$.
In addition, while it would be possible to create every necessary operation required by the BG recursion using finite fields by composing existing TensorFlow primitives, this introduces a significant overhead relative to the actual complexity of the operations.
To mitigate this, we have developed custom kernels (for CPU and GPU) for the subset of operations for which this overhead is greater.
These correspond to the (double and single batched) tensor contraction over an arbitrary number of indices:
\begin{align}
  z_{ni}^{k} = x_{nij\dots} y_{n}^{j\dots k} \\
  z_{ni}^{k} = x_{nij\dots} y^{j\dots k},
\end{align}
as well as the finite field division $\frac{x_{n}}{y}$, with $x,y \in \mathbb{F}_p$ and which is implemented as a multiplication by the inverse of $y$ using the extended Euclidean algorithm.

In the equations above, $n$ corresponds to the event dimension (the batch size of the operation).
These operations can deal with tensors of up to rank 4 with 32-bit indexing.
The implementation can be straightforwardly extended to larger tensors, but it is currently not necessary.
All arithmetic is performed in 64-bit to ensure that no operation can overflow in a single step.
In order to preserve exactitude within $\mathbb{F}_p$, intermediate operations are reduced modulo $p$ (the size of the field).

Parallelisation is performed strictly along the event dimension and tasks that do not depend on the number of events, or that are not performance-critical, are left to the host code.
For example, the recursion algorithm is prepared entirely on the host. Likewise, the input momenta and polarisation states are also prepared on the host and rely on CPU dependencies \texttt{lips}, \texttt{syngular} \cite{DeLaurentis:2023qhd}, and \texttt{Singular} \cite{DGPS}.
While in principle it might be beneficial to offload some other stages to the hardware accelerator, the added complexity does not justify the marginal performance gain at this stage.

The choice of CUDA~\cite{cuda} for the kernels and TensorFlow and python for the overall framework has been driven primarily by the availability of these tools (and compilers) on the computer resources we have access to.
We have deliberately minimised reliance on platform-specific features in order to keep this proof-of-concept project flexible enough for future extensions.
This ensures that kernels can be ported to alternative platforms (\emph{e.g.}~OpenCL) and that the wrapper code can, if necessary, be adapted to other frameworks such as PyTorch~\cite{NEURIPS2019_9015} or C++.

Our approach prioritises flexibility and modularity over performance, with the aim of creating a library easy to use, develop, and extend.
True performance bottlenecks will only emerge once the library is integrated into larger frameworks, and excessive early optimisation might hinder future development.
A limitation of this approach is that, while very recent GPUs can natively operate with up to 128-bit integer (easily enabling for finite fields of size up to $2^{63}$), they are not widespread enough.
As a consequence, we limit the size to $p < 2^{31}$, and we choose as a default $p=2^{31}-19$.
Another crucial aspect of GPU computing is memory management,
but as will be shown in Section~\ref{sec:benchmark}, our implementation scales well without the need for extensive optimisation in this area.
We have thus decided to leave this for future development, as this is not currently a bottleneck for our project.

While the main objective of this work is computing scattering amplitudes on GPUs, the library developed in the context of this work is written such that it can be used standalone for calculations involving finite fields.
The use of python and TensorFlow as the glue language for this project makes the library specially well-suited for machine learning applications or scenarios requiring differentiability~\cite{Heimel:2024wph}.
In such cases, however, further extensions of the library should be developed in order to include GPU kernels for the gradient computations whenever necessary.

The documentation for this library can be found in the following URL: 
\begin{center}
  \url{https://amps-gpu.github.io/BG-Trees/}
\end{center}

\subsection{Validations}

To validate the implementation, we have performed several checks. Most
of them are implemented as continuous integration tests in the
\texttt{tests} directory.

Starting from internal consistency checks, we verify the
implementation of custom GPU kernels for finite field operations
(\texttt{test\_finite\_gpufields.py}). The metric and vertices are
verified to yield the same results on CPU and GPU
(\texttt{test\_metric\_and\_vertices.py}). For the states
(a.k.a.~polarisation vectors), we verify the completeness relations of
Eqs.~\eqref{eq:completess_relation_d_dim} and
\eqref{eq:completess_relation_d_dim_split} in $d=\{4,5,6,7,8,9,10\}$
(\texttt{test\_states.py}). For the $d$-dimensional phase space point
generation, we verify on-shell relations and momentum conservation
(\texttt{tests/test\_phase\_space.py}).

At the level of the amplitudes, we verify that the Ward identity is
satisfied, that results in four dimensions reproduce known analytic
expressions, and that $d$-dimensional results match against an
external library (\texttt{test\_currents.py}) for both finite fields and double precision floating point complex numbers.  The Maximally Helicity
Violating (MHV) amplitudes are checked against Parke-Taylor formula
\cite{Parke:1986gb}
\begin{align}
 \mathcal{M}_n(1^-,2^-,3^+, \ldots, n^+) = \ri g^{n-2} \frac{\langle 12 \rangle^4}{\langle 12 \rangle \langle 23 \rangle \cdots \langle n1 \rangle} ,
\end{align}
for a variety of multiplicities $n$ (implemented in the CI as
$n=5$). We also verify NMHV amplitudes against known analytic
expressions at $n=6$ and $n=7$, of which we retain the alternating
NMHV amplitude at six-point as a representative.
Lastly, our test suite shows how to reproduce the output of {\sc
  Caravel}~\cite{Abreu:2020xvt} when the momenta live beyond four
dimensions.  In order to perform this validation and to allow for
a one-to-one comparison, we have allowed for a modification of the metric
in our code
to match {\sc Caravel}'s alternating metric, $\text{diag}(1, -1, 1,
-1, 1, -1, \dots)$.  We generate $d$-dimensional phase space points
(incorporating the metric change) and then pass the resulting
$d$-momenta to {\sc Caravel}.  By using its \texttt{Forest} debug
feature, we can obtain values for the states in {\sc Caravel}'s
convention which then can be used as the input of our BG
recursion. This allowed us to match the result at the level of the
color-ordered amplitude, up to a power of $2$, which is easily accounted
for by $\mathbb{Q}$-reconstruction from the $\mathbb{F}_p$ value.

\subsection{Benchmarks}
\label{sec:benchmark}

We end this section with a benchmarking of the library presented in this article by running it on different hardware platforms.

\begin{figure}[t]
  \centering
  \begin{subfigure}{0.49\textwidth}
    \includegraphics[width=\textwidth]{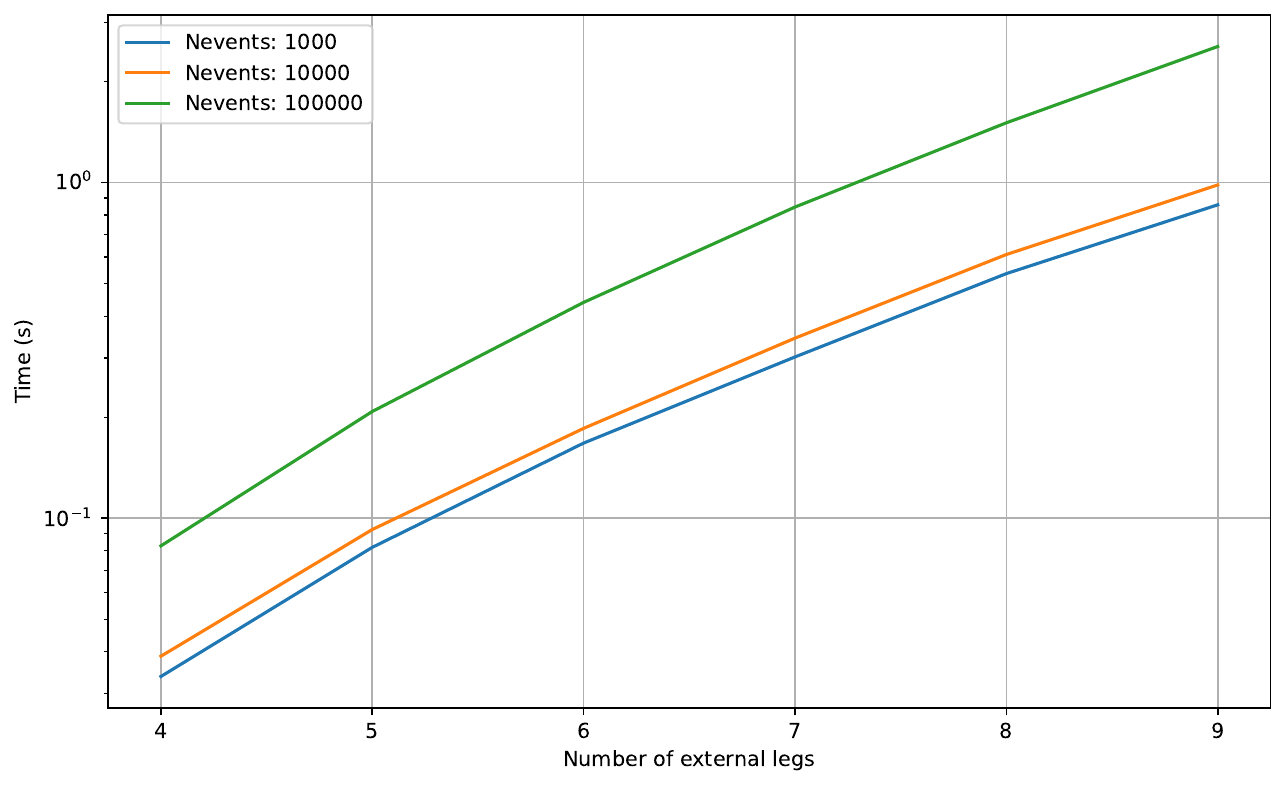}
    \subcaption{Scaling per leg. \label{fig:scaling_per_leg}}
  \end{subfigure}
  \hfill
  \begin{subfigure}{0.49\textwidth}
    \includegraphics[width=\textwidth]{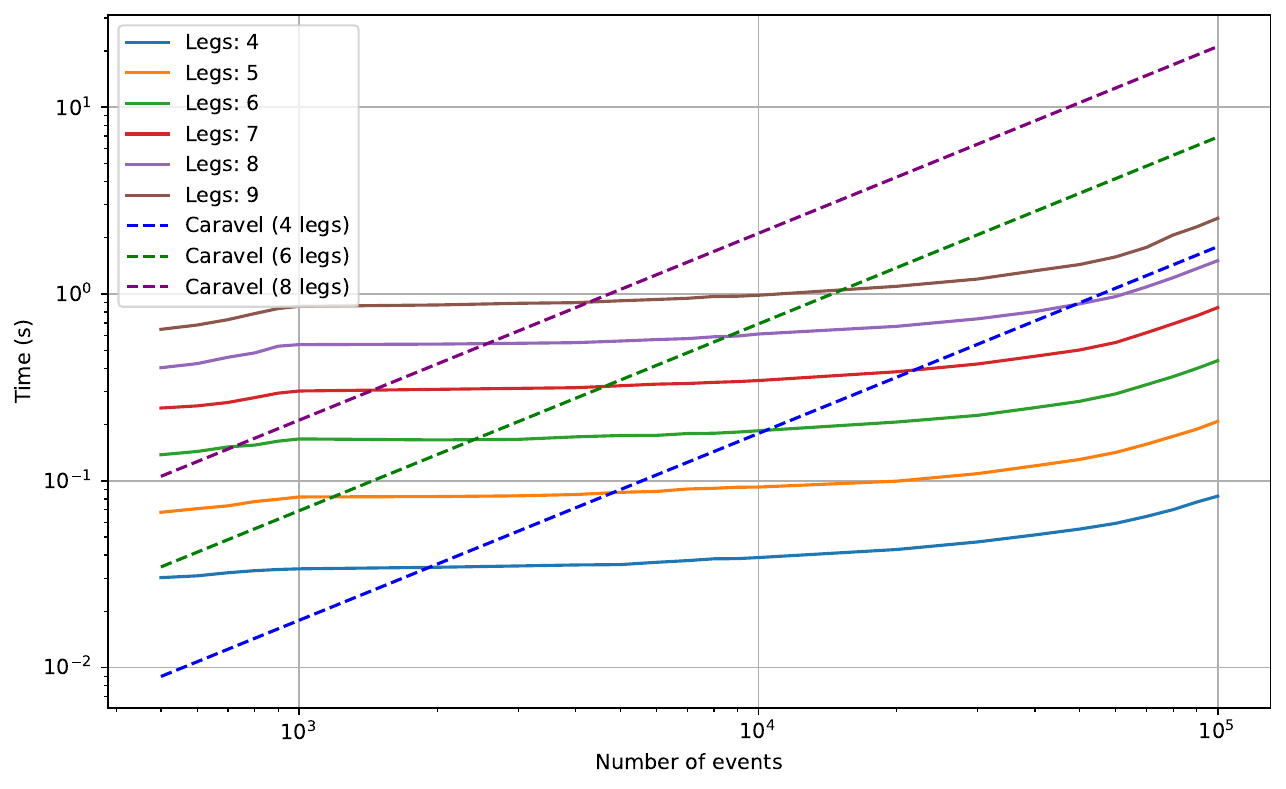}
    \subcaption{Scaling per events.\label{fig:scaling_per_event}}
  \end{subfigure}
  \caption{Total time it takes to compute the amplitude for the given number of phase space points in 8 dimensions for amplitudes with increasing number of external legs for one helicity configuration. All runs are done in GPU. For comparison, the time taken by Caravel running in CPU for 4, 6 and 8 legs (and 8 dimensions) is also shown.}
  \label{fig:perleg_perevent}
\end{figure}

In Fig.~\ref{fig:perleg_perevent}, we show the scaling of the calculation as a function of the number of external legs (Fig.~\ref{fig:scaling_per_leg}) and number of events (Fig.~\ref{fig:scaling_per_event}).
While running on graphics cards speeds up the calculation considerably,
we can see in Fig.~\ref{fig:scaling_per_leg} that the exponential scaling in the number of legs is maintained and independent of the number of events being run.
This can be easily understood, as the change of hardware has no effect on the complexity of the calculation itself.
Instead, the benefits of running on GPU can be immediately seen when we study how the computational cost scales with the number of events.
In Fig.~\ref{fig:scaling_per_event}, the time it takes to finish the computation is shown as a function of the number of events. This time is kept approximately constant as we are filling up the computational capabilities of the GPU up to a threshold. From that threshold onwards the scaling becomes linear.
Note that the threshold does not depend on the number of legs but only on the size of the input array of phase space points. Increasing the complexity of the calculation shifts the curve, but it does not affect the general behavior.

\begin{figure}[t]
  \centering
  \begin{subfigure}{0.49\textwidth}
    \includegraphics[width=\textwidth]{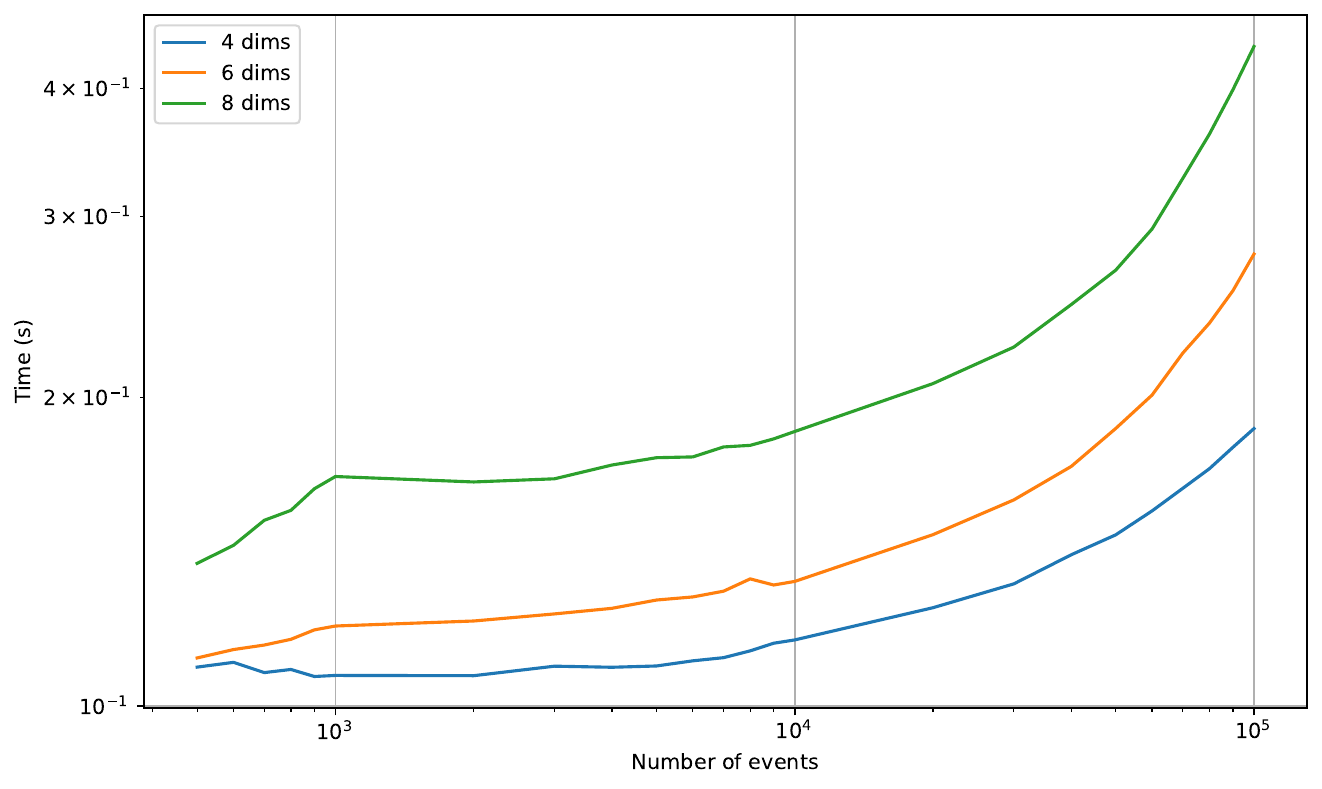}
    \subcaption{6 external particles}
  \end{subfigure}
  \hfill
  \begin{subfigure}{0.49\textwidth}
    \includegraphics[width=\textwidth]{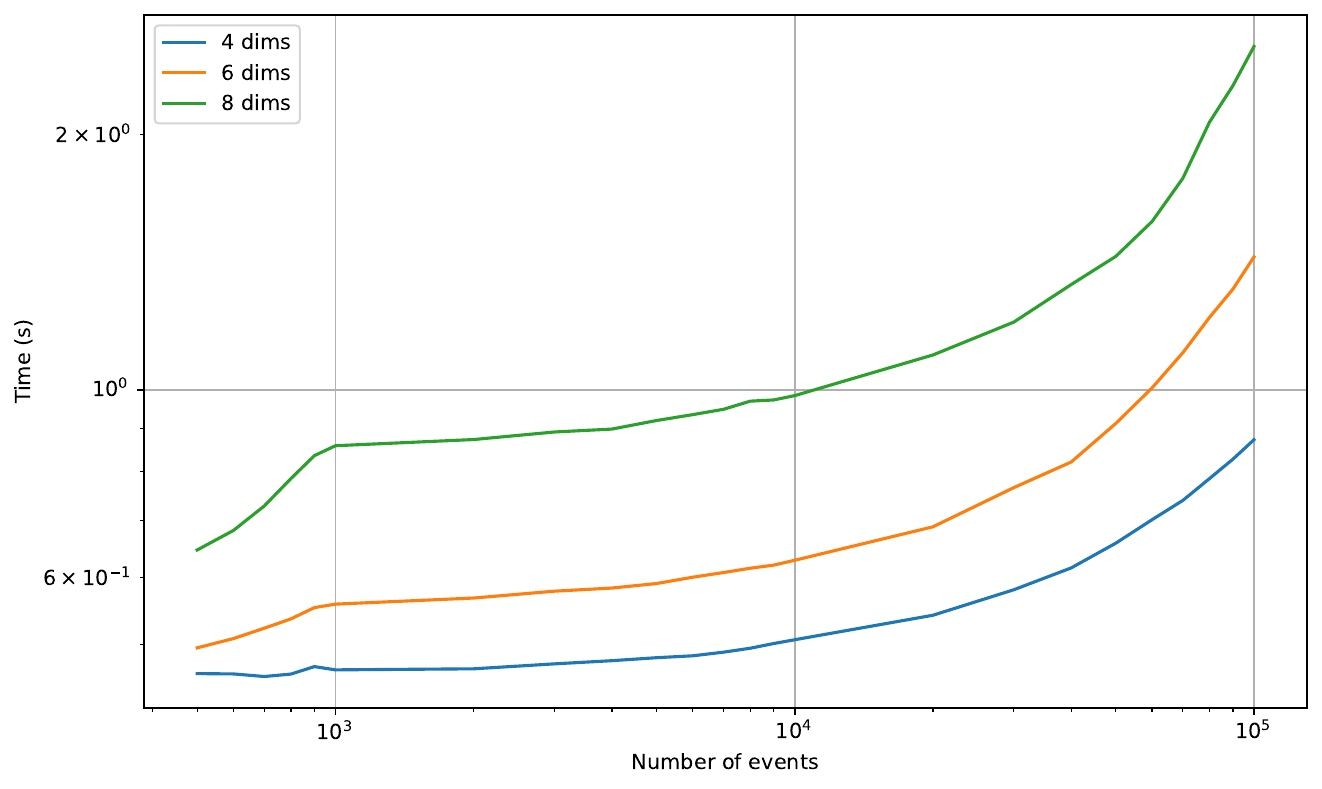}
    \subcaption{9 external particles}
  \end{subfigure}
  \caption{Total time it takes to compute the amplitude for the given number of phase space points for different dimensions for one helicity configuration. We show the 6 (left) and 9 (right) particles cases.}
  \label{fig:perdimension}
\end{figure}

In Fig.~\ref{fig:perdimension}, instead we benchmark the framework for a different number of dimensions for two choices of the number of external particles (6 and 9).
While the computational cost grows with the number of dimensions, the growth is slower compared to the growth with the number of external particles, meaning that extending the computation to an arbitrary number of dimensions can be achieved.

\begin{figure}[t]
  \centering
  \includegraphics[width=0.49\textwidth]{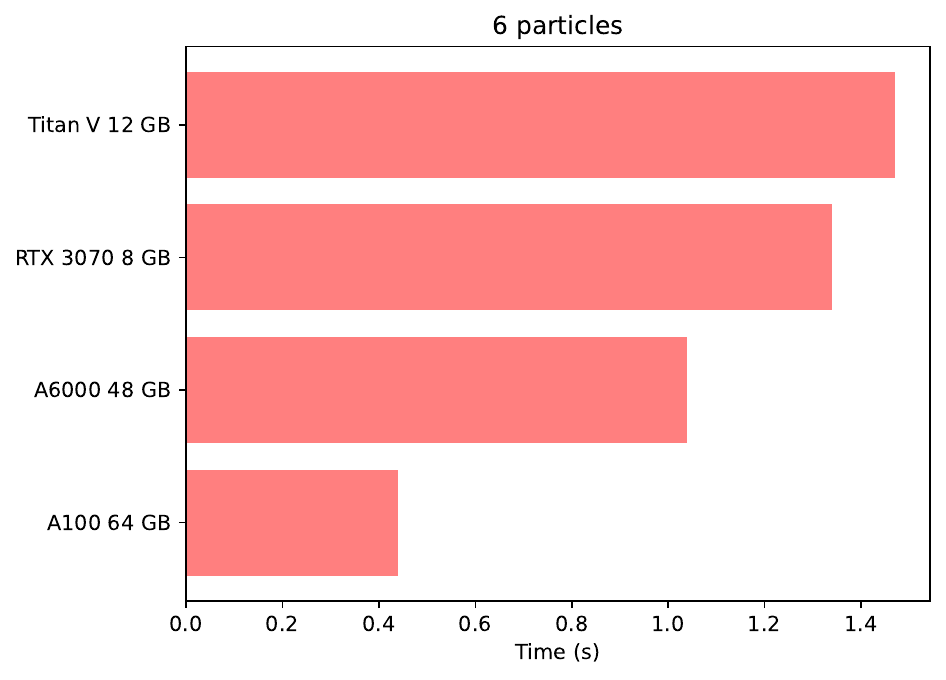}
  \includegraphics[width=0.49\textwidth]{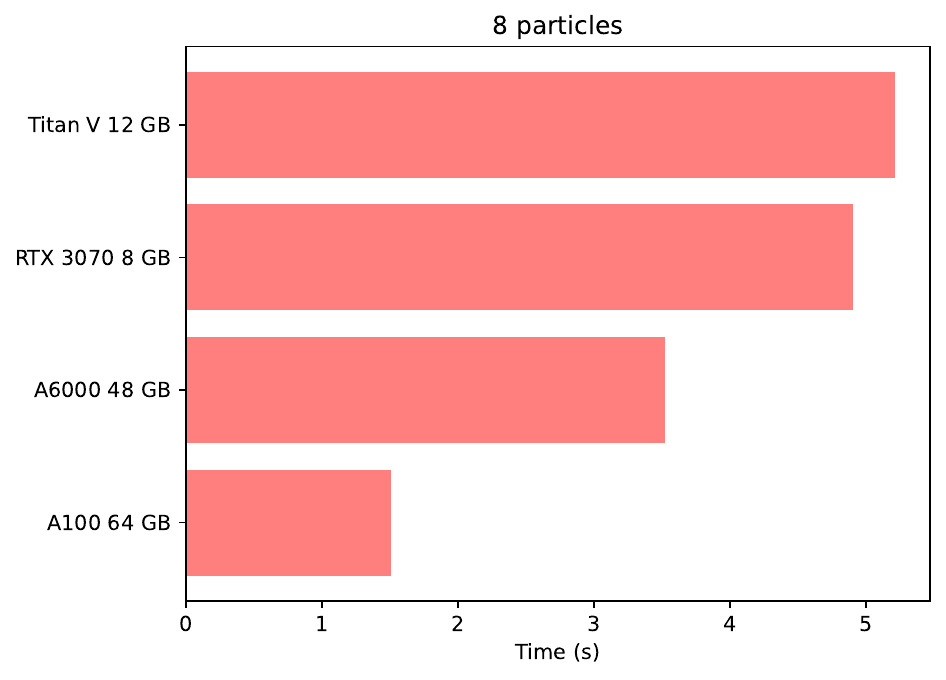}
  \caption{Total time it takes to compute the amplitude for $10^{5}$ events, 6 and 8 external particles, 4 dimensions for different devices. Note that each device is running in a computer with different capabilities, however we observe a general (expected) positive correlation between the memory of the device and the speed, regardless of the number of external particles being considered.}
  \label{fig:perdevice}
\end{figure}

In the last of these comparisons, in Fig.~\ref{fig:perdevice}, we show the behaviour of the code on different devices.
We choose a representative set of graphics cards that can be found both on laptops and home desktop computers as well as enterprise-grade GPUs.
Note that this cannot be considered a fair comparison between the devices since the rest of the hardware in which they are running should also be considered and (for practical reasons) is wildly different in each case.
On the other hand, the similarity in runtime across different hardware is very good news, as it means the choice of running on GPU does not necessarily require costly CPU resources to use it adequately. 

In all plots in Figs.~\ref{fig:perleg_perevent},~\ref{fig:perdimension}, and~\ref{fig:perdevice} we observe also a notable offset even for a small number of events.
This is partly due to the prioritisation of flexibility over performance.
However, in cases where performance is critical, while the overall scaling is expected to remain unchanged, the computing time can be reduced by ahead-of-time compilation (\emph{e.g.}~CUDA kernels for a set number of legs, dimension, etc).
The overhead caused by the choice of python as the language to glue the components together can also be trivially eliminated by changing those components to a compiled language such as C, Rust or Fortran.

\begin{figure}[t]
  \centering 
  \includegraphics[width=0.8\textwidth]{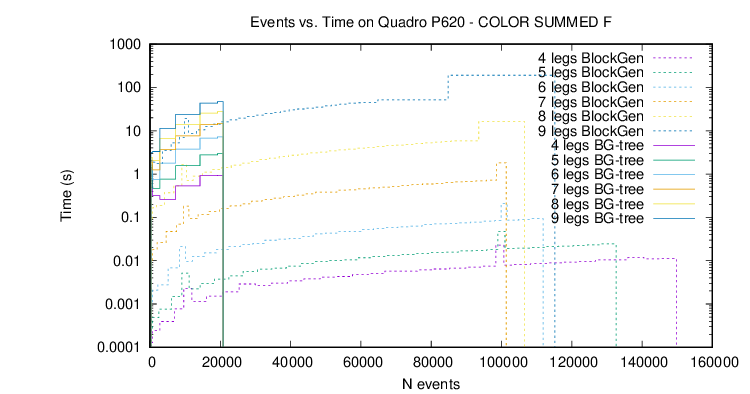}
  \caption{Comparison between the framework introduced in this work and \texttt{BlockGen}~\cite{Bothmann:2021nch,Bothmann:2023gew} for 6 and 9 external particles. For this benchmark both calculations are run in 4 dimensions. For this work, solid line corresponds to finite fields while the dashed line corresponds to executing our code using double precision complex numbers.}
  \label{fig:bgdifference}
\end{figure} 

We conclude with a comparison to another GPU implementation of BG recursion named \texttt{BlockGen} and used in the code \texttt{Pepper}~\cite{Bothmann:2021nch,Bothmann:2023gew}.
While the scopes of this work and \texttt{BlockGen}~\cite{Bothmann:2021nch} are quite different, it is illustrative to benchmark both methods.
The target of \texttt{Pepper} is the calculation of physical observables in a way similar to that of other event generators like Sherpa~\cite{Bothmann:2019yzt}, to achieve this, it uses highly optimised CUDA kernels for double-precision and a four-dimensional momenta phase space.
Our focus instead is on computing amplitudes in an arbitrary number of dimensions using finite fields, with a very flexible code base which can serve other purposes, and so only the most critical components are compiled ahead of time.
With these differences in mind, a benchmark for 6 and 9 particles in the final state is presented in Fig.~\ref{fig:bgdifference}.
As expected, the just-in-time compilation nature of our framework introduces an overhead that requires a large number of events to overcome.
Eventually, the difference in performance between both approaches stabilises with \texttt{BlockGen} being about one order of magnitude faster when comparing against our default finite fields approach.
The difference is slightly smaller when we run our code in double precision mode.
In this benchmark both codes are running with the same number of dimensions (4) and using the leading-color approximation for the calculation.
Note however that the intended use of the different codes is quite different: a custom type ( finite fields) vs.~double precision floats, and dimension-independence vs.~specific for 4 dimensions.

\section{Conclusion}
\label{sec:concl}

In this work we have developed a tree-level matrix element generator
and evaluator based on the Berends-Giele recursion algorithm and
optimised for execution on GPUs. Our primary objective was to
demonstrate the advantages of hardware acceleration for computations
extending beyond four dimensions and employing number types beyond
traditional floating-point arithmetic, specifically finite fields.
Our results show that hardware acceleration, which has so far been
developed primarily for Monte Carlo integration, can indeed be
extended to support higher-order calculations, where products of tree
amplitudes in higher dimensions yield loop-level integrands.

Given the proof of concept nature of this work, we have explicitly
prioritised flexibility over performance. So while we have developed
dedicated kernels for the most costly operations, the more complex
stages of the algorithm are performed in higher-level languages so
that they can be easily modified and possibly improved.

A natural extension of this work is to cover the full Standard Model
by handling also quarks and the electroweak part. This would enable
the integration with existing tools, such as \textsc{Caravel} or Monte
Carlo generators.

\section*{Acknowledgements}

The authors thank Fernando Febres Cordero and Ben Page for useful discussions.
GDL's work is supported in part by the U.K.\ Royal Society through
Grant URF\textbackslash R1\textbackslash 20109.
MP acknowledges support by the German Research Foundation (DFG) through the Research Training Group RTG2044.

\bibliographystyle{utphys.bst}
\bibliography{gpu-BG}
\end{document}

%% file: macros.tex
\newcommand{\ri}{\mathrm i}

\let\origfootnote\footnote
\renewcommand{\footnote}[1]{\kern.06em\origfootnote{#1}}
\newcommand{\punctfootnote}[1]{\kern-.06em\origfootnote{#1}}

\def\be{\begin{equation}}
\def\ee{\end{equation}}

\newcommand{\newc}{\newcommand}
\newc{\bi}{\begin{itemize}}
\newc{\ei}{\end{itemize}}
\newc{\benu}{\begin{enumerate}}
\newc{\eenu}{\end{enumerate}}
\newc{\bc}{\begin{center}}
\newc{\ec}{\end{center}}
\newc{\bfig}{\begin{figure}}
\newc{\efig}{\end{figure}}
\newc{\qbar}{\bar{q}}
\newc{\go}{\tilde{g}}
\newc{\PB}{\textsc{Powheg-Box}}

\newcommand{\madgraph}{{\sc\small MadGraph5\_aMC@NLO}\xspace}

\newcolumntype{.}{D{.}{.}{-1}}
\newcolumntype{d}[1]{D{.}{.}{#1}}

\colorlet{tableoverheadcolor}{gray!37.5}
\colorlet{tableheadcolor}{gray!25}
\colorlet{tablerowcolor}{gray!12.5}

\newlength{\width}
\newlength{\height}

\def\draftdate{\relax}
\def\mda{\relax}
\def\mua{\relax}
\def\mla{\relax}
\def\draft{
\def\thtystars{******************************}
\def\sixtystars{\thtystars\thtystars}
\typeout{}
\typeout{\sixtystars**}
\typeout{* Draft mode!
         For final version remove \protect\draft\space in source file *}
\typeout{\sixtystars**}
\typeout{}
\def\draftdate{\today}
\def\mua{\marginpar[\boldmath\hfil$\uparrow$]%
                   {\boldmath$\uparrow$\hfil}\color{black}%
                    \typeout{marginpar: $\uparrow$}\ignorespaces}
\def\mda{\color{red}\marginpar[\boldmath\hfil$\downarrow$]%
                   {\boldmath$\downarrow$\hfil}%
                    \typeout{marginpar: $\downarrow$}\ignorespaces}
\def\mla{\marginpar[\boldmath\hfil$\rightarrow$]%
                   {\boldmath$\leftarrow $\hfil}%
                    \typeout{marginpar: $\leftrightarrow$}\ignorespaces}
\def\Mua{\marginpar[\boldmath\hfil$\Uparrow$]%
                   {\boldmath$\Uparrow$\hfil}\color{black}%
                    \typeout{marginpar: $\uparrow$}\ignorespaces}
\def\Mda{\color{red}\marginpar[\boldmath\hfil$\Downarrow$]%
                   {\boldmath$\Downarrow$\hfil}%
                    \typeout{marginpar: $\downarrow$}\ignorespaces}
\def\Mla{\marginpar[\boldmath\hfil\textcolor{red}{$\Rightarrow$}]%
                   {\boldmath\textcolor{red}{$\Leftarrow $}\hfil}%
                    \typeout{marginpar: $\leftrightarrow$}\ignorespaces}
\overfullrule 5pt
\oddsidemargin 15mm
\marginparwidth 29mm
}

\newcolumntype{C}{>{\centering\arraybackslash}p{0.105\textwidth}}
